\documentstyle[aps,epsfig,prl]{revtex}

\def\comment#1{}
\def\cm#1{}

\everymath={\displaystyle}

\newcommand{\LL}{\Lambda}

\newcommand{\N}{N_c}

\newcommand{\dslash}{\partial\!\!\!/}

\newcommand{\gd}[1]{\gamma_{#1}}

\newcommand{\ld}[1]{\lambda_{#1}}

\newcommand{\p}{\partial}
\newcommand{\f}[2]{\frac{#1}{#2}}
\newcommand{\tr}{{\rm tr}}
\newcommand{\be}{\begin{equation}}
\newcommand{\ee}{\end{equation}}
\newcommand{\beqn}{\begin{eqnarray}}
\newcommand{\eeqn}{\end{eqnarray}}
\newcommand{\Tr}{{\rm Tr}}

\newcommand{\s}{\sigma}

\newcommand{\SigM}{{M}}

\newcommand{\calL}{\mbox{${\cal L}$}}

\newcommand{\psibar}{\bar{\psi}}

\newcommand{\betacrit}{\beta^{\rm cr}}
\def\cm#1{}

\def\comment#1{}
\title{Nonlinear sigma model
approach for chiral
fluctuations \\ and symmetry breakdown
in Nambu--Jona-Lasinio model.}
\author{Egor Babaev\footnote {
email: egor@teorfys.uu.se
Tel: +46-18-391902, Fax +46-18-533180
}}
\address{
Institute for Theoretical Physics, Uppsala University
Box 803, S-75108 Uppsala, Sweden }
\begin{document}
\maketitle
\begin{abstract}
This paper is organized in two parts. We start 
with an observation that the recent claim
that the chiral symmetry in NJL
model is necessarily 
restored by violent chiral fluctuations at $N_c=3$
 (H. Kleinert and B. Van den Bossche, Phys. Lett. B 474
336 (2000) ) appears to be incorrect since 
the critical stiffness of the effective
nonlinear sigma model used in the above reference is
not an universal quantity in 3+1 - dimensions.
In the second part we discuss 
a modified NJL model, where the critical 
stiffness is expressed via an additional cutoff parameter. This
model displays a symmetry breakdown 
and also under certain
conditions the chiral fluctuations give rise to a phase analogous to 
pseudogap phase of strong-coupling and low carrier 
density superconductors.
\end{abstract}
\section{Introduction}
Many concepts of particle physics have a close
relation to superconductivity, for example
Nambu--Jona-Lasinio model \cite{NJLM}-\cite{ht}
was proposed in analogy to BCS theory and is considered to be a low-energy effective
theory of QCD. Recently there was made a substantial
progress in the theory of superconductivity
in systems with strong attraction and low carrier
density.
Namely
it was observed that away from
the limits of infinitesimally weak coupling strength
or very high carrier density, the BCS-like mean-field theories
are qualitatively wrong and these systems
possess along with a superconductive phase
an additional phase where
there exist Cooper pairs but no symmetry is broken
due to violent  fluctuations
({\it pseudogap phase}). What may
be regarded as an indication of the
importance of this concept to particle physics is
that recently we have found a formation
of the pseudogap phase
due to dynamic quantum fluctuations at low $N$
in the chiral Gross-Neveu model \cite{GNM} in $2+\epsilon $
dimensions \cite{gn1}. 
The chiral GN model at low $N$ exhibit two 
phase transition at two characteristic values 
of renormalized coupling constant $g$. 
At a certain value $g^*$ a gap modulus 
forms locally, but there exists as well 
another characteristic 
coupling value $g_{KT} > g^*$ where the
system undergoes a Kosterlitz-Thouless transition into 
a quasiordered state. 
The region between $g_{KT}$ and $g^*$ is 
analogous to the pseudogap phase in superconductors.
At very large $N$, $g_{KT}$ merges with $g^*$ thus recovering 
BCS-like scenario for the phase transition in the chiral GN model.
Recently,  an attempt was made 
\cite{kb} to generalize this result to the NJL model that lead 
the  authors of \cite{kb} to the
conclusion that at $N_c=3$
the NJL model does not display a spontaneous breakdown
of the chiral symmetry due to  fluctuations.
 The paper \cite{kb}
is at the moment a subject of numerous
controversial discussions. Below we present a "no-go" result
that  one can not prove in principle in analogy to \cite{gn1}
the necessary
restoration of the chiral symmetry at low $N_c$ in the NJL model.
As it is discussed in details below, the reason is the nonuniversality of the
critical stiffness of the effective $O(4)$ - nonlinear sigma model (NLSM)
in $3+1$-dimensions.  This results in
the fact that the $3+1$-dimensional
theory possesses an additional nonspecified
fitting parameter that should be fixed from phenomenological
considerations. 
This is in contrast to the chiral GN model in $2+\epsilon$-dimensions
where 
one can prove that the model does not exhibit 
a quasi-long range order when the number of field 
components  $N$ drops below $8$ \cite{gn1}.

In the last section 
we discuss possibility
of formation of a phase analogous to the pseudogap phase
in a modified NJL model
taking special care of the existence of an
additional cutoff  parameter. 

As it was mentioned above the subject of the
discussion, which is the possibility of the 
restoration of chiral symmetry
by directional fluctuations in a degenerate valley 
of an effective potential while preserving
nonzero gap modulus locally, is closely related
to the pseudogap phenomena in superconductors.
Separation of the  temperatures of the pair formation and
of the onset  of the phase coherence (pair condensation)
in strong-coupling superconductors
is in fact known already for many years (Crossover from BCS
superconductivity to -- Bose-Einstein Condensation (BEC)
of tightly bound fermion pairs)
\cite{Le,N}.
Intensive theoretical studies
of these phenomena in the past several years
(see for example \cite{sc}-\cite{pist}),
were sparked by experimental
results on
underdoped (low carrier density)
cuprates that display  a "gap-like" (pseudogap) feature
{\it above} $T_c$. 
The pseudogap  disappears only at the substantially higher
temperature $T^*$.
There is experimental evidence that
this phenomenon in high-$T_c$ superconductors
may be connected with precritical pairing
fluctuations above $T_c$.

Due to intimate relation of
many problems in particle
physics to superconductivity
it seems to be natural to guess that
pseudogap may become a
fruitful concept in high energy physics too.
We  start with a brief introduction to this phenomena in superconductors
and discuss its possible implications for QCD.

It is a well known fact that the 
 BCS theory describes perfectly metallic superconductors.
However it failed to describe even qualitatively
superconductivity in underdoped High-$T_c$ compounds.
One of the most exotic properties of the later materials
is the existence of a  pseudogap  in the spectrum of the normal state
well above critical temperature. From experimental point
of view this  manifests itself
as an essential suppression of low frequency spectral weight thus
being in contrast to exactly zero spectral weight in the case of
superconductive gap. Moreover spectroscopy experiments
show that  superconductive gap evolves
smoothly by magnitude and wave vector dependence to the pseudogap
in the normal state. Except for it NMR and tunneling
 experiments indicate
existence of incoherent Cooper pairs well above $T_c$. In principle it
is easy to guess what is hidden behind this circumstances,
 and why BCS theory is incapable to describe it.
Let us imagine for a moment that we were able
to bind electrons in Cooper pairs infinitely tightly -
obviously this implies that characteristic temperature
of thermal pair decomposition will be also
infinitely high, however this does not imply that     the
long-range order will survive at infinitely high temperatures.
As it was first observed in \cite{N}, the
long-range order will be destroyed in a similar way as it happens say in
superfluid ${}^4He$ i.e. tightly bound Cooper pairs,
at certain temperature will acquire a nonzero momentum and
thus we will have a gas of tightly bound
Cooper pairs but no macroscopic occupation of
zero momentum level ${\bf q} =0$
and with it no long-range order. Thus a phase diagram
of a strong coupling superconductor has three regions:
\begin{itemize}
\item
Superconductive phase where there are condensed fermion pairs,
\item
{\it Pseudogap} phase where there exist
fermion pairs but there is no condensate
and with it there is no symmetry breakdown and no superconductivity,
\item
 Normal phase with thermally decomposed Cooper pairs.
\end{itemize}
Of course, the existence  of bound pairs above critical temperature
will result in deviations from Fermi-liquid behavior that
makes pseudogap phase to be a very interesting object of
study.
In order to describe superconductivity in a such system
theory should incorporate pairs with
nonzero momentum. Thus { \it the BCS scenario
is invalid for description of spontaneous symmetry breakdown
in a system
with strong attractive interaction or low carrier density}
(see \cite{N}- \cite{pist}). So in principle
in a  strong-coupling superconductor the onset of a long range order has
nothing to do with a pair formation transition.
Existence of the paired fermions is a necessary but not sufficient condition for
symmetry breakdown.
BCS limit is a rather exotic case
of infinitesimally weak coupling strength and high carrier density
when the disappearance of superconductivity
can  be {\it approximately} described  as  a pairbreaking transition.
Strong-coupling limit is another exotic
case when temperatures of the pair decomposition and
symmetry breakdown can be arbitrarily separated. There is nothing surprising
in it: formally in the case of Bose condensation of ${}^4He$ we can also
introduce a characteristic
temperature of thermal decomposition of a ${} He$  atom,
however this would not mean that this temperature will be somehow
related  to  the temperature of the Bose condensation of the
gas of these atoms.

Schematic phase diagram of a superconductor
is shown in Fig ~1.
%
\begin{figure}[tb]
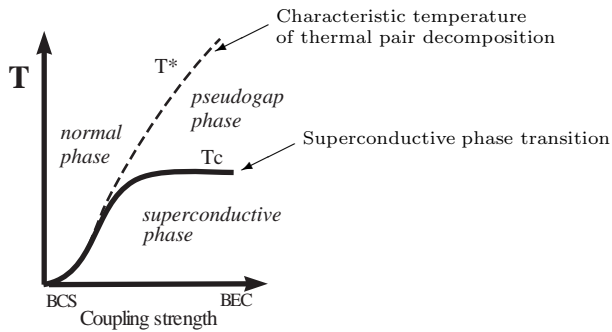

\input Phases.tps
\caption[]{Schematic phase diagram of a superconductor
with arbitrary coupling strength. In the strong coupling limit,
temperature of superconductive phase transition
tends to a plateau value corresponding
to temperature of Bose condensation of the gas of tightly
bound fermion pairs, whereas characteristic temperature of
thermal pair decomposition grows monotonously
as a function of the coupling strength.}
\label{phases.tps}\end{figure}
One can obtain a
pseudogap phase starting from
BCS Hamiltonian. This was first done in a pioneering work
by  Nozieres and Schmitt-Rink \cite{N}
and in a formalism of functional integral by Sa de Melo, Randeria and Engelbrecht
\cite{R}.
In order to study behavior of $T_c$ one should solve
a set of  number and gap equations including fluctuation
corrections.
In the BCS limit $T_c$ is not affected substantially by the gaussian
corrections and superconductive transition
can be described by a 
mean-field theory and correspondingly $T_c \approx T^*$
\footnote{As it was first discussed
in the sixties \cite{asl}, even in the BCS superconductors there is a narrow
fluctuation region near $T_c$
that gives rise e.g. to a so-called paraconductivity effect.
In particle physics, it was pointed on this
phenomenon by Hatsuda and Kunihiro \cite{ht2,ht}. }.
In the opposite limit numerical solution, and analytic
perturbative treatment \cite{N,R} shows
that the temperature of the superconductive phase transition tends
to a constant value
$ T_c= \left[{n}/{2\zeta(3/2)}\right]^{2/3}{\pi}/{m}$
 that does not depend on the coupling strength
and is equal to the temperature of the condensation of the
ideal Bose gas of particles of mass $2m$ and density $n/2$,
where $m$ and $n$ are the mass and the 
density of electrons. Qualitatively the same 
behavior is reproduced in a nonperturbative 
 nonlinear sigma model
approach analogous to discussed in this article \cite{sc}.
One can observe that although there
is no superconductivity in the pseudogap phase
it exhibits reach exotic non-Fermi-liquid behavior due to
pairing correlations  that makes it as
interesting object of theoretical and
experimental study as superconductive phase itself.
\comment{ In particular,
along with specific heat, optical conductivity
and tunneling experiments there are following
circumstances observed in the pseudogap phase:
In experiments on YBCO a significant suppression of
in-plane conductivity $\sigma_{ab}(\omega)$
was observed at frequencies below 500 ${\rm cm}^{-1}$  beginning
at temperatures much above $T_c$.
Experiments on underdoped samples revealed
deviations from the linear resistivity law. In particular
$\sigma_{ab}(\omega=0;T)$
increases slightly with decreasing $T$
below a certain temperature.
NMR and neutrons observations
show that below temperatures $T^*$ much higher than $T_c$,
 spin susceptibility starts decreasing.
}
%
One should stress however that the
term "pseudogap", originated in early experimental
papers, may seem somewhat misleading since even though
a substantial depletion of low-frequency spectral weight is
observed in this region experimentally - there is
{\it no true gap} in the spectrum.
\section{Chiral fluctuations in the the NJL model at zero temperature}
As it was mentioned above, recently it was made an attempt
\cite{kb} of generalization to the NJL model
the nonlinear-sigma approach  for description of chiral
fluctuations proposed in \cite{gn1,sc}.
The authors \cite{kb}
claimed that at $N_c=3$ the NJL
model does not display spontaneous  symmetry breakdown
due to chiral fluctuations.
We show below that
NLSM approach does not allow
 to prove that the chiral symmetry is
always restored  
by  fluctuations in the NJL model at $N_c=3$.
Below we also discuss difference  with the chiral GN model 
where  NLSM approach 
allows one to reach a similar conclusion at low $N$  \cite{gn1}.

The Lagrangian of the NJL model reads \cite{NJLM}:
\be
\calL=\psibar
i\dslash
\psi+\f{g_0}{2N_c}\left[
\left(
\psibar\psi
\right)^2+\left(
\psibar\ld{a}i\gd{5}\psi
\right)^2
\right].
\label{NJLModel}
\ee
The three $2\times2$-dimensional
matrices $ \ld{a}/2$, generate the fundamental representation
of flavor $SU(2)$, and are normalized by
$\tr (\ld{a}\ld{b})=2\delta_{ab}$.
One can introduce Hubbard - Stratonovich fields
$ \s$ and $\pi_a$:
\be
\calL=\psibar\left(
i\dslash-\s-i\gd{5}\ld{a}\pi_a
\right)
\psi-\f{\N}{2g_0}\left(
\s^2+\pi_a^2
\right).
\label{hsnjl}
\ee
After integrating out quark fields, following to
a standard mean-field variation procedure
one can choose pseudoscalar solution $\pi_a$ to be vanishing
and scalar solution $\sigma\equiv M$ to be given by a gap equation:
\be
\f{1}{g_0}=i(\tr_f 1)(\tr_{\gamma} 1)\int
\f{d^Dp}{(2\pi)^D}\f{1}{p^2-M^2}
\label{gap1}
\ee
The momentum integral is regularized by means of a cutoff $\Lambda$.
\comment{
Mean-field  $N_c\rightarrow \infty$
treatment gives an effective action
\be
 \Gamma (\rho)=- \Omega [\Delta v( \rho )+v_0]
\label{EffectivePotential}
\ee
where $\Omega$ is the spacetime volume, and $v_0$
is energy density of the symmetric state,
whereas
\beqn
 &&\Delta v(\rho)=\f{N_c}{2}\Bigg\{
\f{1}{g_0}\rho^2-\f{2}{(2\pi)^2}\Bigg[
\f{\rho^2\Lambda^2}{2}
+\f{\Lambda^4}{2}\ln\left(
1+\f{\rho^2}{\Lambda^2}
\right)\nonumber\\
&&\mbox{}-\f{\rho^4}{2}\ln\left(
1+\f{\Lambda^2}{\rho^2}
\right)
\Bigg]
\Bigg\}
\label{lambdapotential}
\eeqn
the mean-field condensation energy at {\it constant}
$  \sigma ^2+\pi_a^2\equiv  \rho ^2$.
The momentum integral is regularized by means of a cutoff
$\Lambda$.
The condensation energy is extremal
at
$ \rho = \SigM$ which solves the {\em gap equation\/}
\beqn
 \f{1}{g_0}&=&\f{2}{(2\pi)^2}
\left[
\Lambda^2- \SigM^2\ln\left(
1+\f{\Lambda^2}{\SigM^2}
\right)
\right].
\label{lambdagap}
\eeqn    }
The constituent quark mass $\SigM$
in the limit $\N\rightarrow \infty$
is analogous to superconductive gap
in the BCS limit of the theory of superconductivity.

At finite $N_c$ one can study fluctuations
around the saddle point solution.
The quadratic terms of expansion
around  the saddle point are:
\begin{eqnarray}
{\cal A}_0[\s',\pi'] = \f{1}{2}\!\int\!\!d^4q\!\left[
\left(\!\!
\begin{array}{c}
\pi'_a(q)\\
\s'(q)
\end{array}
\!\!\right)^T\!\!
\left(\!\!
\begin{array}{cc}
G_{\pi}^{-1}&0\\
0&G_{\s}^{-1}
\end{array}
\!\!\right)\!\!
\left(\!\!
\begin{array}{c}
\pi'_a(-q)\\
\s'(-q)
\end{array}
\!\!\right)
\right],
\label{ao}\end{eqnarray}
where
$(\s',\pi'_a)\equiv(\s- \SigM,\pi_a)$
and $G_{\s,\pi}^{-1}$
are the inverse bosonic propagators.
\comment{
\begin{equation}
G_{\s}^{-1} =
\N\left[ 2\times2^{D/2} \int
\f{d^4p_E}{(2\pi)^4}\f{(p_E^2+p_Eq_E - M^2)}
{(p_E^2 + M^2)[(p_E+q_E)^2 +M^2]}
- \f{1}{g_0}\right];
G_{\pi}^{-1} =
\N\left[ 2\times2^{D/2} \int
\f{d^4p_E}{(2\pi)^4}\f{(p_E^2+p_Eq_E + M^2)}
{(p_E^2 + M^2)[(p_E+q_E)^2 +M^2]}
- \f{1}{g_0}\right].
\end{equation}
In the above expression one should introduce
a momentum cutoff $\Lambda_2$.} Implementing  a momentum
cutoff $\Lambda$, we can write
$G_{\pi,\sigma}^{-1}$ for small $q_E$ as:
\beqn
\!\!\!\!\!\!G_{\pi}^{-1}\!\approx\!\!-\f{\N}{(2\pi)^2}
\!\left[
\ln\left(
1\!+\!\f{\Lambda^2}{\SigM^2}
\right)
\!-\!\f{\Lambda^2}{\Lambda^2\!+\!\SigM^2}\right]
\!q_E^2\!\equiv\!\!-Z(\SigM/\Lambda)q_E^2; \ \ \ \ \ \ \ \ \
 G_{ \sigma }^{-1}\!\approx\!\!
-Z(\SigM/\Lambda)(q_E^2+4\SigM^2).
\label{SigPropStiffLambda}
\eeqn
In analogy to $3D \ XY$-model approach to
strong-coupling superconductivity \cite{sc}
the authors of \cite{kb} introduced
a unit vector field
$n_i\equiv (n_0,n_a)\equiv(\s,\pi_a)/ \rho $
and set up an effective nonlinear sigma-model
\be
{\cal A}_0[n_i]= \f{\beta}{2}\int d^4x
[\partial n_i(x)]^2.
\label{@prop}
\ee
The prefactor $\beta=M^2 Z(M/\Lambda)$,
 that follows from (\ref{ao}),
(\ref{SigPropStiffLambda}) 
is playing the role of a stiffness
of the unit field fluctuations.
%
%
%

Now let us observe
that from the arguments given in
\cite{kb} it does not follow that NJL model necessarily
remains in a chirally symmetric phase
at $N_c=3$.
At first, in contrast to discussed in \cite{gn1}
 $2+\epsilon$-dimensional case,
one unfortunately, can not made any similar calculations in a
closed form in $3+1$-dimensions because this
theory is not renormalizable. 
It was already observed in \cite{cut}-\cite{bub} that
cutoff of meson loops can not be set equal to cutoff
for quark loops and thus the $1/N_c$ corrected theory
\cite{bub} possesses two
independent  parameters that may be adjusted at will. 
We present another arguments 
of a different nature
rooted in a nonuniversality of a 
critical stiffness of a NLSM in four dimensions,
that does not allow 
in the  framework 
of the NLSM approach 
to reach the conclusion of \cite{kb}.
Our observation 
also applies to NLSM
description of precritical fluctuations in general systems. 
It also allows us to show that the discussed below additional
cutoff can not be related to the inverse coherence length 
of the radial fluctuations in the effective potential as it was suggested in 
\cite{prl,kb}.

Basically the  authors of \cite{kb} by deriving
$G_{\pi, \sigma}$ have extracted two 
characteristics from the initial system:
stiffness of the phase fluctuations in the degenerate minimum
of the effective potential
and the mass of the radial fluctuation. However knowledge of these
characteristics does not allow in principle to judge 
if directional fluctuations will destroy
long range order or system will possess  a BCS-like
phase transition.
The reason is that a critical stiffness of the nonlinear sigma
model is not an universal quantity in $3+1$-dimensions.
So in principle knowledge of the stiffness of NJL model 
is not sufficient for finding the
{\it position} of the phase transition in the effective
nonlinear sigma model.
The situation is just like in a Heisenberg
magnet where the critical temperature depends
on the stiffness along with lattice spacing and lattice
structure. Thus if one is given only a
stiffness coefficient one can not determine
temperature of the phase transition\footnote{Due to this
reason one can not refer to lattice simulation for 
finding the value of the critical stiffiness as it was done in \cite{kb,prl}
since implicitly these numerical values contain information 
of lattice structure and are not universal.}.
The situation is
in contrast to 2D case when a position of a
KT-transition can be deduced from the stiffness coefficient \cite{bkt}.
In two dimensions the critical stiffness of O(2) nonlinear 
sigma model is an universal quantity and is given by 
$\beta_{KT}=2/\pi$ \cite{bkt}, so comparing 
it with the stiffness coefficient 
derived from the initial theory 
(phase stiffness of the chiral GN model in $D=2$ is
$\beta = N/4\pi$ )
one can judge if the 
system has enough stiffen phase in order to 
preserve quasi-long range order 
as we have shown in \cite{gn1}
\footnote{There is a misleading 
statement about three dimensional case in \cite{gn1}.}. I.e. one can determine
the number of field components N 
 that is needed to remain below the
position of Kosterliz-Thouless transition.
This is 
in contrast to discussed here  $D=3+1$ case.


Let us first recall a procedure how one 
can express a critical stiffness
of the O(4)-nonlinear sigma model via 
an additional parameter:
one can relax constrain $n_i^2 =1$
and introduce an extra integration over the
lagrange multiplier $\lambda$ rewriting (\ref{@prop}) as:
%
$(\beta/2) \int d^4x
\left\{ [\partial n_i(x)]^2+ \lambda \left[ n_i^2(x)-1\right] \right\}$.
%
Integration out the $n_i(x)$-fields, yelds:
\be
{\cal A}_0[\lambda]=-\beta\int d^4x  \f{\lambda(x)}{2}+\f{N_n}{2}\Tr\ln\left[
-\p^2+\lambda(x)
\right],
\label{newaction}
\ee
where $N_n$ is the number of components of $n_i(x)$, and $\Tr$ denotes the
functional  trace.
This yields a gap equation:
\be
\beta=N_n\int \f{d^4k}{(2\pi)^4}\f{1}{k^2+\lambda} .
\label{@secge}\ee
The model has  a phase transition at a  critical stiffness
that depends on an unspecified  additional cutoff parameter that 
should be applied to the gap equation:
\be
\betacrit=N_n\int \f{d^4k}{(2\pi)^4}\f{1}{k^2}.
\label{CriticalStiff}
\ee
For example in the case of magnets the additional cutoff
needed in (\ref{CriticalStiff})
is naturally related to the lattice spacing.
In the paper \cite{prl} it was proposed a
criterion that states that one can relate the inverse
coherence length extracted from  radial fluctuations in an
effective potential of an initial theory to the cutoff 
in the integral (\ref{CriticalStiff}) so that all the parameters
in the theory would be expressed from quantities
derived from an initial model and thus this 
modified model would possess an universal 
critical stiffens.
However there is no reason
for relating the cutoff needed in (\ref{CriticalStiff})
to the coherence length of the modulus fluctuations 
 and moreover we show that 
this procedure leads in general to unphysical consequences.
It was supposed in \cite{prl} 
that relation of coherence length to cutoff
in the equation (\ref{CriticalStiff})
yields an universal criterion 
for judgement of 
nature of symmetry breakdown in 
general physical systems.
There is a simple counterexample:
in the case of  a
strong-coupling superconductor, the 
 characteristic nonlinear sigma model
that describes 
fluctuations in a degenerate valley 
of the effective potential is a 3D XY-model.  In
the continuous case it is  a free field
theory and has no phase transition at all.
The phase transition appears only in the lattice theory
and of course its temperature
depends on the lattice spacing.
With increasing coupling strength the 
low-temperature phase stiffness of the effective 3D XY model tends
to a plateau value
$J=n/4m$, where $n$ and $m$ are density and mass
of fermions \cite{sc}. Whereas the temperature of the phase transition
of the 3D XY-model is 
\be
T_c^{3D XY} \propto  \frac{n}{m} a,
\label{xy}
\ee
 where
$a$ is the lattice spacing. 
To be careful one should remark that accurate analysis shows
that 
a strong coupling superconductor possesses two characteristic length 
scales: size of the Cooper pairs  that tends to zero with increasing 
coupling strength 
and a coherence length that tends to infinity with increasing coupling
strength as the system evolves towards a weakly nonideal gas
of true composite bosons \cite{R,pist}.
At first if one relates the constant $a$ in (\ref{xy}) to the size of 
the Cooper pairs 
following to the arguments of \cite{kb}
one would come to an incorrect conclusion of absence of the
superconductivity in strong-coupling superconductors
in the way similar as the authors
of \cite{kb} came to a conclusion 
of inexistence of symmetry 
breakdown
in  the NJL model.
This is in a direct
contradiction with 
behavior of the strong coupling superconductors 
(see references menitioned in the Introduction).
Second, if one attempts to relate $a$ in (\ref{xy}) 
to the second length scale
of the theory - namely true coherence length, that 
tends to infinity with increasing coupling strength, 
then one would come to a qualitatively incorrect conclusion too
\cite{rem}. Thus the existence of an universal NLSM-based
fluctuations criterion \cite{prl} appears to be incorrect.

So in general the nonlinear sigma model approach 
for precritical fluctuations possesses an additional 
fitting parameter which is the
cutoff in the gap equation (\ref{CriticalStiff})
that can not be related to inverse coherence length 
extracted 
from radial fluctuations in an effective potential.
Thus within the NLSM approach  one can 
not proove if the NJL model displays
necessarily the directional fluctuations driven restoration of the chiral 
symmetry at low $N_c$.
\section{Chiral fluctuations at finite temperature and a modified NJL
model with a pseudogap}
The authors of \cite{kb} employed NLSM 
arguments in attempt to show that the NJL model can not serve 
for the study of the chiral symmetry breakdown. We have shown 
above that this conclusion appears to be incorrect
since the critical stiffness
in 3+1-dimensions is not an universal quantity and one 
has an additional fitting parameter.
This is an inborn feature of the discussed NLSM approach
 in 3+1 dimensions
(compare with the mentioned above
cutoffs 
discussions in nonrenormalizable models in a different approach
\cite{cut}-\cite{bub}, and also \cite{bl}).
The above circumstance allows one to fix the critical 
stiffness from phenomenological considerations.
However,  we argue below that, what is missed 
in \cite{kb} is that, in principle, the low-$N_c$
fluctuation instabilities, when properly treated, 
have a clear physical meaning.
Moreover we argue that 
 one can employ a NLSM 
for descriptions of the chiral fluctuations,
providing that special care is taken of
the additional cutoff parameter.
It was indeed already discussed in literature that at finite temperatures
the chiral phase transition should be accompanied by 
developed  fluctuations (\cite{ht,ht2} and references therein).
We argue that this 
process at low $N_c$ should give  rise to a 
phase analogous to the pseudogap phase that may be conveniently 
described within a nonlinear sigma model approach. There are
indeed other ways to describe these phenomena \cite{H}.
 However the NLSM
approach  seems to be especially convenient in the case
of a nonrenormalizable theory. The descrition of 
the two-step chiral phase transition and appearence of the 
intermediate phase requires to study the system at the next-to-mean-field
level. Unfortunately the NJL model is not renormalizable 
and does not allow to make any conclusions about 
importance of fluctuations in a closed form \cite{bub}.
On the other hand a pseudogap phase 
is a general feature of fermi system with attraction.
The NLSM construction discussed below, due to
its nonperturbative nature can not be 
regarded as a regular approximation but may be 
considered as a tractable modification of the NJL model that 
has a pseudogap.
One can also find an additional
motivation for  employing these arguments in the fact that 
NLSM allows one
to prove an existence of a phase analogous to pseudogap phase
in the chiral GN model \cite{gn1} which is the closest relative to NJL model.
Also NLSM approach works well for the description of 
precritical fluctuations in superconductors \cite{sc} - where 
the theory is renormalizable and 
essentially the same results can be obtained perturbatively.
We stress that
these phenomena is a general feature 
of any Fermi system with attraction.
Also, to certain extend similar crossovers 
are known in a large variety of condensed matter systems.
In particular, besides  superconductors
we can mention the
exitonic condensate in
semiconductors, the
Josephson junction arrays,
the itinerant and local-momentum theories
of magnetism and  the ferroelectrics.

Let us now consider the chiral fluctuations in NJL model
at finite temperature.
Then
 following standard dimensional reduction 
arguments \cite{Wil}, the chiral fluctuations should be
 described by a  $3D \ O(4)$-sigma model.
Thus one  has 
the following gap equation for the effective NLSM
(i.e. finite temperature analogue of (\ref{@secge})):
\be
\frac{J_T}{T} = N_n \int \frac{d^3 k}{(2\pi)^3} \frac{1}{k^2+\lambda}
\label{gt}
\ee
The temperature of the phase transition of the three dimensional
classical $O(4)$ sigma
model with stiffness $J_T$ is expressed via an additional parameter
${\tilde \Lambda}_T$ needed in (\ref{gt}) as :
\be
T_c =  \frac{\pi^2}{2}\frac{J_T}{{\tilde \Lambda}_T}
\label{tc}
\ee
The stiffness of thermal fluctuations 
$J_T$ can be readily extracted from the NJL model.
At finite temperature the inverse bosonic propagator of the collective
field $\pi$ for small $q$ can be written as:
\begin{eqnarray}
G^{-1}_\pi &= & -2^{D/2} N_c \int \frac{d^3p}{(2\pi)^3} \sum_n
\left[ \frac{T}{(p^2+M^2+\omega_n^2)^2}\right] q^2 =
\nonumber \\
& - &  2^{D/2} N_c \int \frac{d^3 p}{ (2 \pi)^3}
\left[ \frac{1}{8}
\frac{1}{(p^2+M^2)^{3/2}}
\tanh
\left( \frac{\sqrt{p^2+M^2}}{2 T}\right)
-\frac{1}{16 T}\frac{1}{p^2+M^2} \cosh^{-2}
\left(
\frac{\sqrt{p^2+M^2}}{2T}
\right)
\right] q^2 = \nonumber \\
& - &  K (T,\Lambda_T, M, N_c) q^2,
\label{st0}
\end{eqnarray}
where $\Lambda_T$ is a momentum cutoff. 
The propagator (\ref{st0}) renders the 
gradient term that allows one to set up an effective 
classical $3D ~ O(4)$-nonlinear sigma model :
\be
E=\frac{J_T (T, \Lambda_T, M, N_c)}{2} \int d^3 x [\partial n_i (x)]^2,
\label{Hei}
\ee
where 
\be
J_T(T, \LL_T, M, N_c) = K(T, \LL_T, M, N_c) ~M^2(T,\Lambda_T)
\label{st1}
\ee
is the stiffness of the thermal  fluctuations in the degenerate 
valley of the effective potential. The temperature -dependent quark
mass $M$ that enters this expression 
is given by a standard mean-field gap equation that 
also should be regularized with the cutoff $\Lambda_T$:
\be
\f{1}{g_0}=2\times2^{D/2}\sum_n
\int \f{d^3p}{(2\pi)^3}\f{T}{p^2+ M^2 +\omega_n^2}.
\label{gap}
\ee
It can be easily seen that
when we approach the temperature $T^*$
where the mass the $M(T)$ becomes zero, the stiffness
$J(T,\Lambda_T, M, N_c)$ also tends to zero.
Formula (\ref{Hei}) defines a generalized Heisenberg model
with a {\it temperature dependent stiffness coefficient}.
Position of the phase disorder transition in a such 
system should be determined self-consistently by
solving the system of the equations for $T_c$ and $M(T_c)$.
Apparently just like 
in a superconductor
with a pseudogap 
the phase transition in a such system is a competition
between a thermal depletion of the gap modulus (this 
rudely corresponds to thermal pairbreaking in a superconductor)
and a process of thermal excitations of 
the directional  fluctuations in the
degenerate minimum of the effective potential.
"BCS" limit corresponds to the situation when $T^*$ merges with
$T_c$ and it is easily seen that this scenario always holds true at
 $N_c \rightarrow \infty$. I.e. at infinite $N$ the mean-field
theory is always accurate just like BCS theory works well in the
weak coupling superconductors. In the framework 
of this NLSM construction, at low $N_c$ the scenario 
of the phase transition
depends on the  choice of $M(0), \Lambda_T $ 
and ${\tilde \Lambda}_T$, that should be fixed from 
phenomenological considerations.


\section{Conclusion}
In the first part of this
paper we presented a no-go result that
within a framework of the nonlinear sigma model approach one
can not answer the question if the chiral
symmetry in NJL model is always restored by quantum
fluctuations at  $N_c=3$. The reason is 
the nonuniversality
of the critical stiffness of $4 D \ O(4) $ nonlinear sigma model.
This, along with the discussed above observations made in
a framework of a different approach in \cite{bub}
resolves numerous controversial discussions
initiated by a recent paper \cite{kb}
where it was argued that 
there is no spontaneous breakdown of the 
chiral symmetry in the NJL model,
which appears to be incorrect.

In the second part of the paper we discussed a NLSM
approach for precritical fluctuations 
in a modified NJL model where a critical stiffness is expressed via
an additional cutoff parameter that should be fixed 
from phenomenological consideration. We discuss a
formation in the above model 
of  a phase analogous to the pseudogap phase
in strong-coupling and low-carrier density superconductors.
Appearence of this phase  
may accompany the chiral phase transition in  QCD.
Since the precursor pairing fluctuations
is a general feature of any Fermi system with attraction
and moreover it is a dominating region of a phase
diagram
of strong-coupling and low carrier density superconductors,
the interesting question is to what extend the discussed 
above phase
is developed in QCD and color superconductors.
\comment{
Indication of possible importance of the pseudogap
concept in particle physics is the mentioned above existence of this
phenomenon in the chiral Gross-Neveu model at low $N$.
Even though these results can not be directly generalized to
NJL model, one can guess that in analogy
to 3D XY-model approach to strong-coupling
and low carrier density superconductivity, one
can set up a nonlinear 3D O(4)-sigma
model with temperature depended stiffness
coefficient as a toy model for QCD at finite temperatures
that would possess two characteristic temperatures
corresponding to discussed in this paper $T_c$ and $T^*$.
Speaking about the BCS-BEC crossover,
precritical fluctuations and the pseudogap phase,
it should be noted as well that
in some sense similar phenomena are known
in a large variety of condensed matter systems,
in particular, except for superconductors
we can mention itinerant and local-momentum theories
of magnetism, exitonic condensate in
semiconductors, ferroelectrics and Josephson junction arrays.}

\begin{acknowledgments}
The author is grateful to Prof. A. J. Niemi and Dr. V.V. Cheianov
for discussions, to Prof. T. Hatsuda and Prof. D. Blaschke 
for communicating useful references.
\end{acknowledgments}


\begin{thebibliography}{999}
\bibitem{rem}
In fact, in a nonperturbative 
NLSM approach to the  condensation from 
pseudogap phase in strong-coupling superconductors one can reach 
an excellent agreement with the perturbative results \cite{N,R}
if one considers an effective sigma model on the lattice 
with the spacing given by $a \propto 1/n_{b}^{1/3}$,
where $n_b$ is  density of Cooper pairs \cite{sc}. Then 
the model is related to the condensation of 
hard-core composite bosons on a lattice.
\end{thebibliography}
\end{document}